\documentstyle[11pt,adassconf]{article}  

\begin{document}   

%

\paperID{P2-43}


\title{Infrared Imaging Data Reduction Software and Techniques}

%

\author{C.N.\ Sabbey, R.G.\ McMahon, J.R.\ Lewis, \& M.J.\ Irwin}
\affil{Institute of Astronomy, Madingley Road, Cambridge, CB3 0HA, UK}


\contact{Chris Sabbey}
\email{sabbey@ast.cam.ac.uk}

%
%

\paindex{Sabbey, C. N.}
\aindex{Lewis, J. R.}     
\aindex{McMahon, R. G.}     
\aindex{Irwin, M. J.}




\begin{abstract}          

We describe the InfraRed Data Reduction (IRDR) software package, a small
ANSI C library of fast image processing routines for automated pipeline
reduction of infrared (dithered) observations.  We developed the software
to satisfy certain design requirements not met in existing packages (e.g.,
full weight map handling) and to optimize the software for large data sets
(non-interactive tasks that are CPU and disk efficient).  The software
includes stand-alone C programs for tasks such as running sky frame
subtraction with object masking, image registration and coaddition
with weight maps, dither offset measurement using cross-correlation,
and object mask dilation.  Although we currently use the software to
process data taken with CIRSI (a near-IR mosaic imager), the software
is modular and concise and should be easy to adapt/reuse for other work.
IRDR is available from anonymous ftp to ftp.ast.cam.ac.uk in pub/sabbey.

\end{abstract}


\section{Introduction}

The Cambridge Infrared Survey Instrument (CIRSI) is a near-IR mosaic
imager that contains a 2 x 2 array of Rockwell Hawaii I 1024 x 1024
detectors (Beckett et al.\ 1996; Mackay et al.\ 2000).  CIRSI has been in
operation for about two years and we have obtained almost 1 Terabyte of
imaging data.  The uniquely wide field accessible by CIRSI on a 2--4m
class telescope makes it ideal for large-area surveys to moderate depth.
Two such surveys are currently in progress with preliminary results
including the measurement of galaxy clustering at intermediate redshift
(McCarthy et al.\ 2001), and the demonstration of a reddening-independent
quasar selection technique based on combined deep optical and near-IR
color diagrams (Sabbey et al.\ 2001).

However, the CIRSI data reduction poses several challenges.  With the
large data rate (5--10 GB of data taken per night, $\sim 100$ nights per
year, and a significant data backlog currently) the software has to be
very efficient and completely automated.  Also, the software should
handle diverse data sets, from Galactic center observations to very
sparse fields at high Galactic latitude.  We generate wide-field, deep
mosaic images from thousands of individual images taken over many nights.
With the gaps between detectors comparable to detector size, we make
filled mosaic images using the coadded dither sets from different chips
and telescope pointings.  Thus weight maps and accurate astrometry are
crucial and the common simplification of clipping the dither sets to
their intersection region is not appropriate.

Although we use existing software packages when possible (see below),
we decided to write the core image processing routines ourselves to
satisfy certain design requirements.  For example, we wanted a two pass
reduction with object masks derived from the first pass coadded dither
sets, subpixel image registration and coaddition that uses full weight
maps and does not clip images, optimizations for CPU/disk efficiency,
customized artifact cleaning (destriping and defringing), and reusable
tools (from having stand-alone tasks to glue together using a high
level scripting language, to making C library calls, or even extracting
portions of source code).  The basic processing steps, described below,
are: flatfield correction, running sky frame subtraction, dither offsets
measurement, dither set coaddition, and mosaic image creation.


\section{Data Reduction}
\subsection{Flatfield Correction}
Flatfield images are produced by subtracting the stack median of lamp-off
domeflats from the stack median of lamp-on domeflats.  The flatfield
images are divided by the mode of the chip 1 flatfield to produce a gain
map per chip.  Bad pixels are automatically identified in the gain maps
(and set to 0.0) by looking for outliers ($> 5\sigma$ from the median
in 15x15-pixel blocks) or pixels with extremely low or high sensitivity
($> 30$\% from the median gain).  Because bad pixels often occur in
clumps we eliminate bad pixels during image coaddition rather than by
interpolation in an initial cleaning pass.  The data frames are multiplied
by the inverse of the gain map.

The image stacking is done using {\tt cubemean.c}, which calculates the
median, robust standard deviation, or robust mean plane with a choice
of weights (none, scalar, or maps) and image scaling or zero offsets
using the image modes.  This is not a general purpose tool like IRAF's
{\tt imcombine}, but for the specific task of calculating the median plane
was found to be 2.5 times faster (for a stack of 50 of our data frames).
The flatfield image is converted to a gain map and bad pixels identified
using {\tt gainmap.c}.  The data are flatfield corrected using {\tt flat.c}.

\subsection{Running Sky Frame Subtraction}
For each data frame, a sky image is constructed from the robust mean
of the 8 nearest frames in the observation sequence and subtracted
({\tt skyfilter.c}).  Objects detected in the coadded dither sets from
the first pass reduction are masked out during sky frame creation in the
second pass.  The object masks are produced using the checkimage {\tt
OBJECTS} option to SExtractor (Bertin and Arnouts 1996), which produces
a FITS image with non-object pixels set to 0.  This is simpler and more
effective than building masks from a catalog of object coordinates and
shape parameters.  The object regions (detection isophotes) generated by
SExtractor are then expanded by a multiplicative factor of 1.5 (using
{\tt dilate.c}), thereby growing the mask regions for large objects
more than small objects.  The object masks for individual frames are
obtained on the fly using pixel offsets (i.e., the dither offsets)
into the master dither set masks.



Running sky frame subtraction is normally a significant bottleneck in
processing infrared imaging, so optimizations are important.  To do
running sky frame subtraction for a stack of N images, the program {\tt
skyfilter.c} uses a sliding window (circular buffer of image pointers) to
only require N image reads and N image mode calculations.  In contrast,
putting this logic into a script normally involves N $\times$ M image
reads and mode calculations, where M is the width of the sky filter
in frames.  Also, the disk I/O (and storage) for non-coadded data uses
short integers (2 bytes deep), even though most calculations are done
in floating point (4 bytes).  Some calculations work with short integer
data however to allow optimizations.  For example, the almost trivial
distribution sort can be used to obtain an image histogram, sorted
image array, and median value in $O(n)$ time ($\approx 5$ times faster
than running an optimized median routine on our data images).

%
%
%
%

\subsection{Dither Offsets Measurement}
A typical dither sequence consists of nine observations in a $3 \times
3$ grid with offsets in each direction of $\approx 10$ arcsec.  The
approximate dither offsets stored in the FITS header WCS information are
refined using cross-correlation analysis ({\tt offsets.c}).  The non-zero
(object) pixels of the reference frame object mask (SExtractor OBJECTS
image) are stored in a pixel list (x, y, brightness), and this list is
cross-correlated against the object mask images of the following frames
in the dither set.  The SExtractor OBJECTS image conveniently removes
the background (important for cross-correlation methods) and identifies
the object pixels more reliably than a simple thresholding algorithm
(e.g., especially in images with a non-flat background, large noise, and
cosmic rays).  Using an object list in the cross-correlation focuses on
the pixels that contribute to the cross-correlation signal and is faster
than cross-correlating two images.

The cross-correlation technique uses coordinate, magnitude, and shape
information, and was found to be more reliable than matching object
coordinate lists (the improvement was noticed in extreme cases, like
Galactic center images and nearly empty fields with an extended galaxy).
A subpixel offset measurement accuracy of about $0.1$ pixels is obtained
by fitting a parabola to the peak of the cross-correlation image.
In terms of speed, this cross-correlation method was found to be $\approx
10$ times faster (for typical survey data and a relatively large search
box of 100 pixels) than IRAF STSDAS {\tt crosscor}.  Although the success
rate is $\approx 100$\%, failure is indicated by an offset measurement
corresponding exactly to the border of the search area, or a small fraction
of object pixels overlapping in the aligned data images.

\subsection{Dither Set Coaddition}
A weight map is generated on the fly for each data frame with the weight
for pixel $p_{i}$ given by: $w_{i} = g_{i} \cdot t / V$, where $g_{i}$ is the
gain for pixel $p_{i}$ (0.0 for bad pixels), $t$ is the exposure time, and $V$
is the image variance.  The data frames and corresponding weight maps in
the dither set are registered using bi-linear interpolation modified to
account for bad pixels and image weights.  Each output (interpolated) pixel 
value $P$ is calculated from the weighted average of the four overlapping
input pixels $p_{i}$ of the input image:
$$
P = \frac{1}{W} \sum_{i = 1}^{4} a_{i} w_{i} p_{i}, \; \; \; \;
W = \sum_{i = 1}^{4} a_{i} w_{i}
$$
where $a_{i}$ are the pre-calculated fractional areas of overlap of $P$
with $p_{i}$, and $w_{i}$ are the image weights for $p_{i}$.  The weight
maps are registered similarly but using a weighted sum to calculate
the weight $W$ for pixel $P$.  A different registration method that
people sometimes recommended is to replicate each pixel of an image
into N x N pixels and do an integer shift in units of these new pixels.
Although this will approximate the above bi-linear interpolation scheme
as N becomes large, it requires more work for less precision.  We have
not tested higher order interpolators (e.g., Devillard 2000), although
our default approach is fast and reasonable, especially given the low
signal-to-noise ratio of our individual data frames.



The dither frames are combined by calculating the weighted mean pixel
value at each $(x,y)$ position of the dither stack, with pixel values $>
5 \sigma$ from the median at each position rejected.  Images borders are
added during registration to avoid clipping the data to the intersection
of the dither frames.  The standard deviation ($\sigma$) at each position
is calculated from $\sigma = {\rm MAD} / 0.6745$, where MAD is the median
absolute deviation from the median (we do not use a simpler method such
as minmax rejection because we often take averages of small numbers of
values, e.g., 5--9 frames per dither set).  The weight maps are combined by
calculating the sum at each $(x,y)$ position of the stack of weight maps
(for pixels not clipped during coaddition).  The program used for the
above is {\tt dithercubemean.c}.

\subsection{Mosaic Creation}

The current astrometry pipeline is a small Perl script that
runs SExtractor to produce an object catalog for each coadded
dither set, runs APMCAT (a stand-alone C program available from
www.ast.cam.ac.uk/$\sim$apmcat/) to download over the network the APM
sky coordinates of objects in each field of view, then runs IMWCS from
WCSTools (Mink 1999) to calculate the astrometry fit and update the WCS
information in the FITS headers.  The coadded dither sets and weight
maps from different chips, telescope pointings, and nights are then
drizzled onto a wide-field mosaic image using EIS Drizzle (available
from www.eso.org/eis).  The astrometry residuals between the mosaic
image and the APM catalogue show a random error of $\sigma \approx 0.3$
arcsec without significant systematic effects.

\end{document}